\documentclass[12pt]{article}
\usepackage{graphicx}
\usepackage{verbatim}
\usepackage{nowidow}
\usepackage[utf8]{inputenc}
\usepackage[margin=2.5cm]
{geometry}
\usepackage{url}
\usepackage{authblk}
\usepackage{subfig}

\title{Scaling in Words on Twitter} 
\author[1,*]{Eszter Bok\'anyi}
\author[2,3]{D\'aniel Kondor}
\author[1]{G\'abor Vattay}
\affil[1]{E\"otv\"os Lor\'and University, Budapest, Hungary}
\affil[2]{Senseable City Laboratory, MIT, Cambridge MA 02139 USA}
\affil[3]{Singapore-MIT Alliance for Research and Technology, Singapore}
\affil[*]{E-mail: bokanyi@complex.elte.hu}
\date{}                     
\begin{document}

\maketitle

\begin{abstract}

Scaling properties of language are a useful tool for understanding generative processes in texts. We investigate the scaling relations in citywise Twitter corpora coming from the Metropolitan and Micropolitan Statstical Areas of the United States. We observe a slightly superlinear urban scaling with the city population for the total volume of the tweets and words created in a city. We then find that a certain core vocabulary follows the scaling relationship of that of the bulk text, but most words are sensitive to city size, exhibiting a super- or a sublinear urban scaling. For both regimes we can offer a plausible explanation based on the meaning of the words. We also show that the parameters for Zipf's law and Heaps law differ on Twitter from that of other texts, and that the exponent of Zipf's law changes with city size.

\end{abstract}

\section{Introduction}

The recent increase in digitally available language corpora made it possible to extend the traditional linguistic tools to a vast amount of often user-generated texts. Understanding how these corpora differ from traditional texts is crucial in developing computational methods for web search, information retrieval or machine translation \cite{CrystalInternetLinguistics}. The amount of these texts enables the analysis of language on a previously unprecedented scale \cite{Altmann2016StatisticalLinguistics,Gerlach2014ScalingFrequencies,Altmann2009BeyondWords}, including the dynamics, geography and time scale of language change \cite{Goel2016,Goncalves2017TheEnglish}, social media cursing habits \cite{Wang2014CursingTwitter,Gauthier2015TextHabits,Byrne2014SweetSoccer} or dialectal variations \cite{Blodgett2016c}.

From online user activity and content, it is often possible to infer different socio-economic variables on various aggregation scales. Ranging from showing correlation between the main language features on Twitter and several demographic variables \cite{Bokanyi2016}, through predicting heart-disease rates of an area based on its language use \cite{Eichstaedt2015} or relating unemployment to social media content and activity \cite{Llorente2014,bokanyi2017prediction, Pavlicek2015} to forecasting stock market moves from search semantics \cite{Curme2014}, many studies have attempted to connect online media language and metadata to real-world outcomes. Various studies have analyzed spatial variation in the text of OSN messages and its applicability to several different questions, including user localization based on the content of their posts \cite{Cheng2010,Backstrom2008}, empirical analysis of the geographic diffusion of novel words, phrases, trends and topics of interest \cite{travelingtrends,Eisenstein2012}, measuring public mood \cite{Mitchell2013}.

While many of the above cited studies exploit the fact that language use or social media activity varies in space, it is hard to capture the impact of the geographic environment on the used words or concepts. There is a growing literature on how the sheer size of a settlement influences the number of patents, GDP or the total road length driven by universal laws \cite{Bettencourt2007}. These observations led to the establishment of the theory of urban scaling \cite{Bettencourt2010b,Alves2015c, Arcaute2015a,Cottineau2017,Bettencourt2013a,Bettencourt2013d,Gomez-Lievano2012, Gomez-Lievano2016a,Yakubo2014SuperlinearCities}, where scaling laws with city size have been observed in various measures such as economic productivity \cite{Lobo2013}, human interactions \cite{Schlapfer2014b}, urban economic diversification \cite{Strumsky2016}, election data \cite{Bokanyi2018UniversalResults}, building heights \cite{Schlapfer2015UrbanSize}, crime concentration \cite{Oliveira2017,Hanley2016a} or touristic attractiveness \cite{Bojic2016ScalingStates}.

In our paper, we aim to capture the effect of city size on language use via individual urban scaling laws of words. By examining the so-called scaling exponents, we are able to connect geographical size effects to systematic variations in word use frequencies. We show that the sensitivity of words to population size is also reflected in their meaning. We also investigate how social media language and city size affects the parameters of Zipf's law \cite{Zipf1932SelectedLanguage}, and how the exponent of Zipf's law is different from that of the literature value \cite{Zipf1932SelectedLanguage,Takahashi2018AssessingProperties}. We also show that the number of new words needed in longer texts, the Heaps law \cite{Altmann2016StatisticalLinguistics} exhibits a power-law form on Twitter, indicating a decelerating growth of distinct tokens with city size. 

\section{Methods}

\subsection{Twitter and census data}

We use data from the online social network Twitter, which freely provides approximately 1\% of all sent messages via their streaming API. For mobile devices, users have an option to share their exact location along with the Twitter message. Therefore, some messages contain geolocation information in the form of GPS-coordinates. In this study, we analyze 456 millions of these geolocated tweets collected between February 2012 and August 2014 from the area of the United States. We construct a geographically indexed database of these tweets, permitting the efficient analysis of regional features \cite{Dobos2013}. Using the Hierarchical Triangular Mesh scheme for practical geographic indexing, we assigned a US county to each tweet \cite{Szalay2007,Kondor2014}. County borders are obtained from the GAdm database \cite{gadm}. Counties are then aggregated into Metropolitan and Micropolitan Areas using the county to metro area crosswalk file from \cite{CMSsCrosswalk}. Population data for the MSA areas is obtained from \cite{Bureau}.

There are many ways a user can post on Twitter. Because a large amount of the posts come from third-party apps such as Foursquare, we filter the messages according to their URL field. We only leave messages that have either no source URL, or their URL after the \texttt{'https://'} prefix matches one of the following SQL patterns: \texttt{'twit\%'}, \texttt{'tl.gd\%'} or \texttt{'path.com\%'}. These are most likely text messages intended for the original use of Twitter, and where automated texts such as the phrase 'I'm at'  or 'check-in' on Foursquare are left out.

For the tokenization of the Twitter messages, we use the toolkit published on \url{https://github.com/eltevo/twtoolkit}. We leave out words that are less than three characters long, contain numbers or have the same consecutive character more than twice.  We also filter hashtags, characters with high unicode values, usernames and web addresses \cite{Dobos2013}. 

\subsection{Urban scaling}

Most urban socioeconomic indicators follow the certain relation for a certain urban system:
\begin{equation}
Y(N)=Y_0\cdot N^\beta,
\label{eq:scaling}
\end{equation}
\noindent where $Y$ denotes a quantity (economic output, number of patents, crime rate etc.) related to the city, $Y_0$ is a multiplication factor, and $N$ is the size of the city in terms of its population, and $\beta$ denotes a scaling exponent, that captures the dynamics of the change of the quantity $Y$ with city population $N$. $\beta=1$ describes a linear relationship, where the quantity $Y$ is linearly proportional to the population, which is usually associated with individual human needs such as jobs, housing or water consumption. The case $\beta>1$ is called superlinear scaling, and it means that larger cities exhibit disproportionately more of the quantity $Y$ than smaller cities. This type of scaling is usually related to larger cities being disproportionately the centers of innovation and wealth. The opposite case is when $\beta<1$, that is called sublinear scaling, and is usually related to infrastructural quantities such as road network length, where urban agglomeration effects create more efficiency. \cite{Bettencourt2013a}

Here we investigate scaling relations between urban area populations and various measures of Twitter activity and the language on Twitter. When fitting scaling relations on aggregate metrics or on the number of times a certain word appears in a metropolitan area, we always assume that the total number of tweets, or the total number of a certain word $Y_{tot}$ must be conserved in the law. That means that we have only one parameter in our fit, the value of $\beta$, while the multiplication factor $Y_0$ determined by $\beta$ and $Y_{tot}$ as follows:

\[\sum_{i=1}^K Y_0\cdot N_i^\beta = Y_{tot},\]

\noindent where the index $i$ denotes different cities, the total number of cities is $K$, and $N_i$ is the population of the city with index $i$. 

We use the 'Person Model' of Leitao et al. \cite{Leitao2016a}, where this conservation is ensured by the normalization factor, and where the assumption is that out of the total number of $Y_{tot}$ units of output that exists in the whole urban system, the probability $p(j)$ for one person $j$ to obtain one unit of output depends only on the population $N_j$ of the city where person $j$ lives as
\[p(j)=\frac{N_j^{\beta-1}}{Z(\beta)},\]
where $Z(\beta)$ is the normalization constant, i.e. $Z(\beta)=\sum_{j=1}^M N_j^{\beta-1}$, if there are altogether $M$ people in all of the cities. Formally, this model corresponds to a scaling relationship from (\ref{eq:scaling}), where $Y_0=Y_{tot}/Z(\beta)$. But it can also be interpreted as urban scaling being the consequence of the scaling of word choice probabilities for a single person, which has a power-law exponent of $\beta-1$.

To assess the validity of the scaling fits for the words, we confirm nonlinear scaling, if the difference between the likelihoods of a model with a $\beta_W$ (the scaling exponent of the total number of words) and $\beta$ given by the fit is big enough. It means that the difference between the Bayesian Information Criterion (BIC) values of the two models $\Delta BIC = BIC_{\beta=1}-BIC_{\beta\neq 1}$ is sufficiently large \cite{Leitao2016a}: $\Delta BIC >6$. Otherwise, if $\Delta BIC<0$, the linear model fits the scaling better, and between the two values, the fit is inconclusive.

\subsection{Zipf's law}

We use the following form for Zipf's law that is proposed in \cite{FerreriCancho2001}, and that fits the probability distribution of the word frequencies apart from the very rare words:
\[p(f) = C\cdot f^{-\alpha}\mbox{, if }f>f_{min}.\]

We fit the probability distribution of the frequencies using the \texttt{powerlaw} package of Python \cite{Alstott2014}, that uses a Maximum Likelihood method based on the results of \cite{Goldstein2004,Toeplitz2015,Virkar2014}. $f_{min}$ is the frequency for which the power-law fit is the most probable with respect to the Kolmogorov-Smirnov distance \cite{Alstott2014}.

A perhaps more common form of the law connects the rank of a word and its frequency:
\[f(r)=C\cdot r^{-\gamma}.\]
\noindent We use the previous form because the fitting method of \cite{Alstott2014} can only reliably tell the exponent for the tail of a distribution. In the rank-frequency case, the interesting part of the fit would be at the first few ranks, while the most common words are in the tail of the $p(f)$ distribution.

The two formulations can be easily transformed into each other (see \cite{FerreriCancho2001}, which gives us
\[\alpha=\frac{1}{\gamma}+1.\]
\noindent This enables us to compare our result to several others in the literature.

\section{Results and discussion}

\subsection{Scaling of aggregate metrics}

\begin{figure}[h!]
\begin{center}
\includegraphics[width=0.6\textwidth]{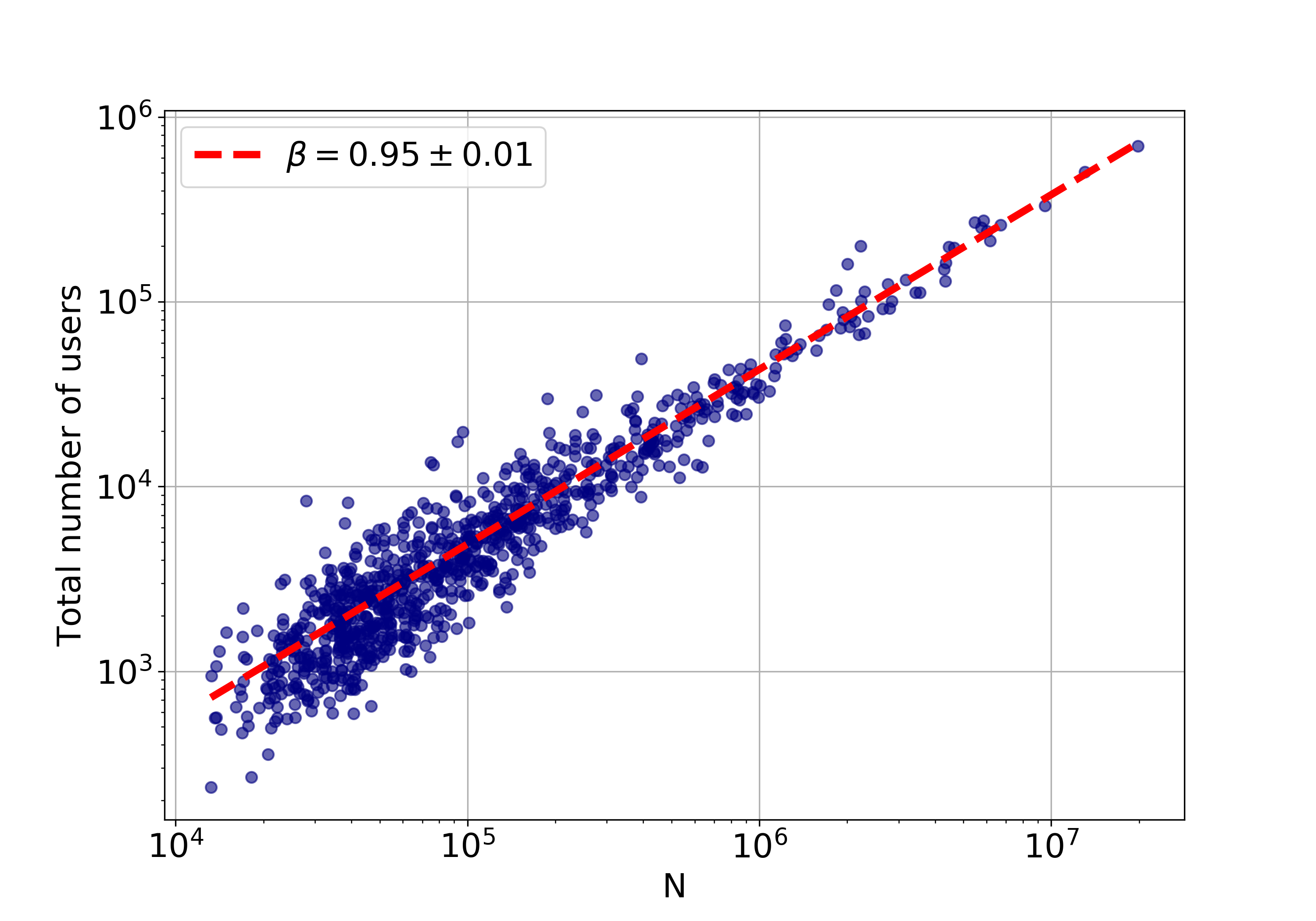}
\caption{Scaling of the number of distinct users who sent a geolocated message with city population. Each point represents an MSA, the fitted line is the best MLE fit for the Person Model of \cite{Leitao2016a}.}
\label{fig:usernum}
\end{center}
\end{figure}

\begin{figure}[p]
\begin{center}
\includegraphics[width=0.6\textwidth]{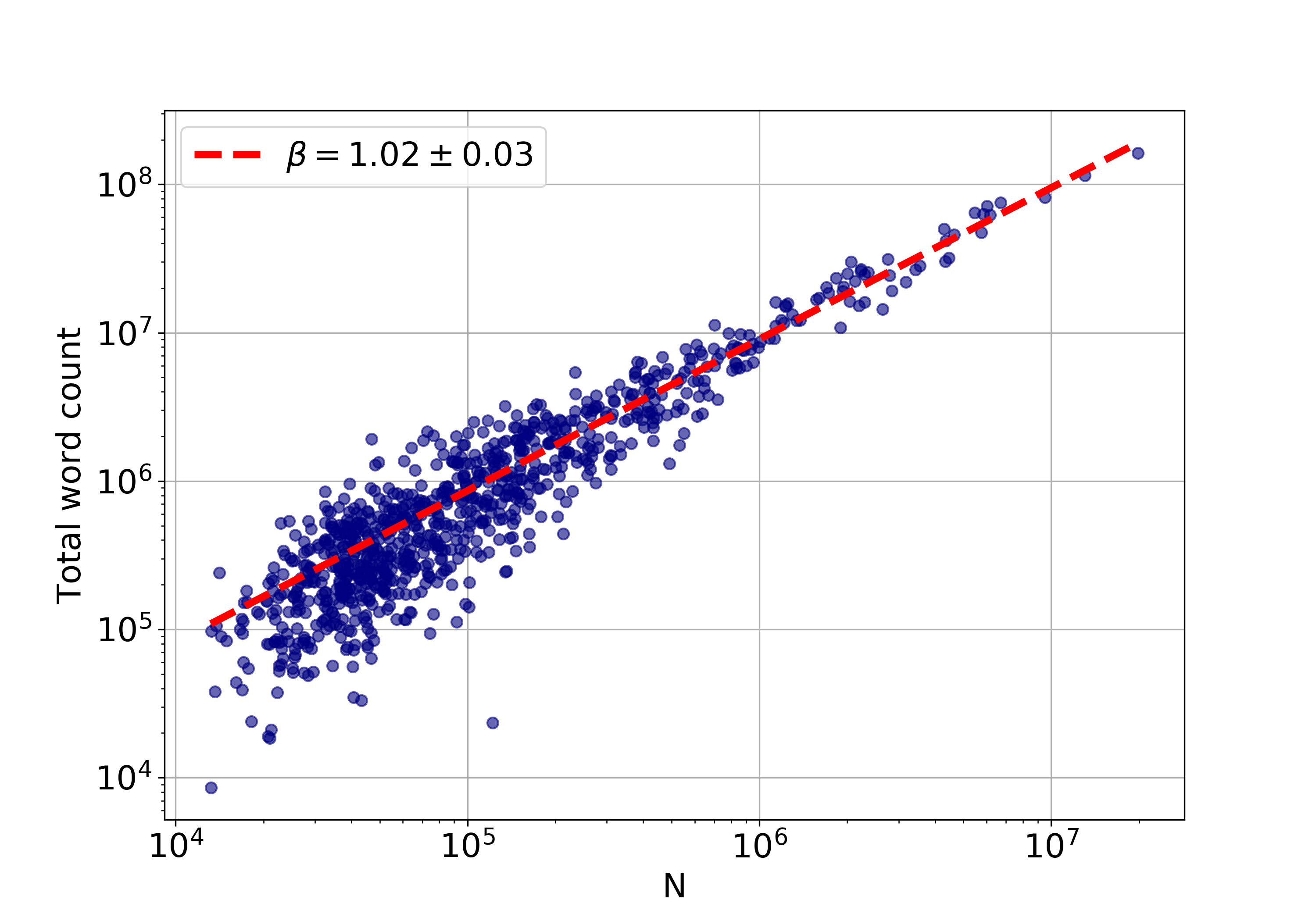}
\caption{Scaling of the total number of words with city population. Each point represents an MSA, the fitted line is the best MLE fit for the Person Model of \cite{Leitao2016a}.}
\label{fig:wordcnt}
\end{center}
\end{figure}

\begin{figure}[p]
\begin{center}
\includegraphics[width=0.6\textwidth]{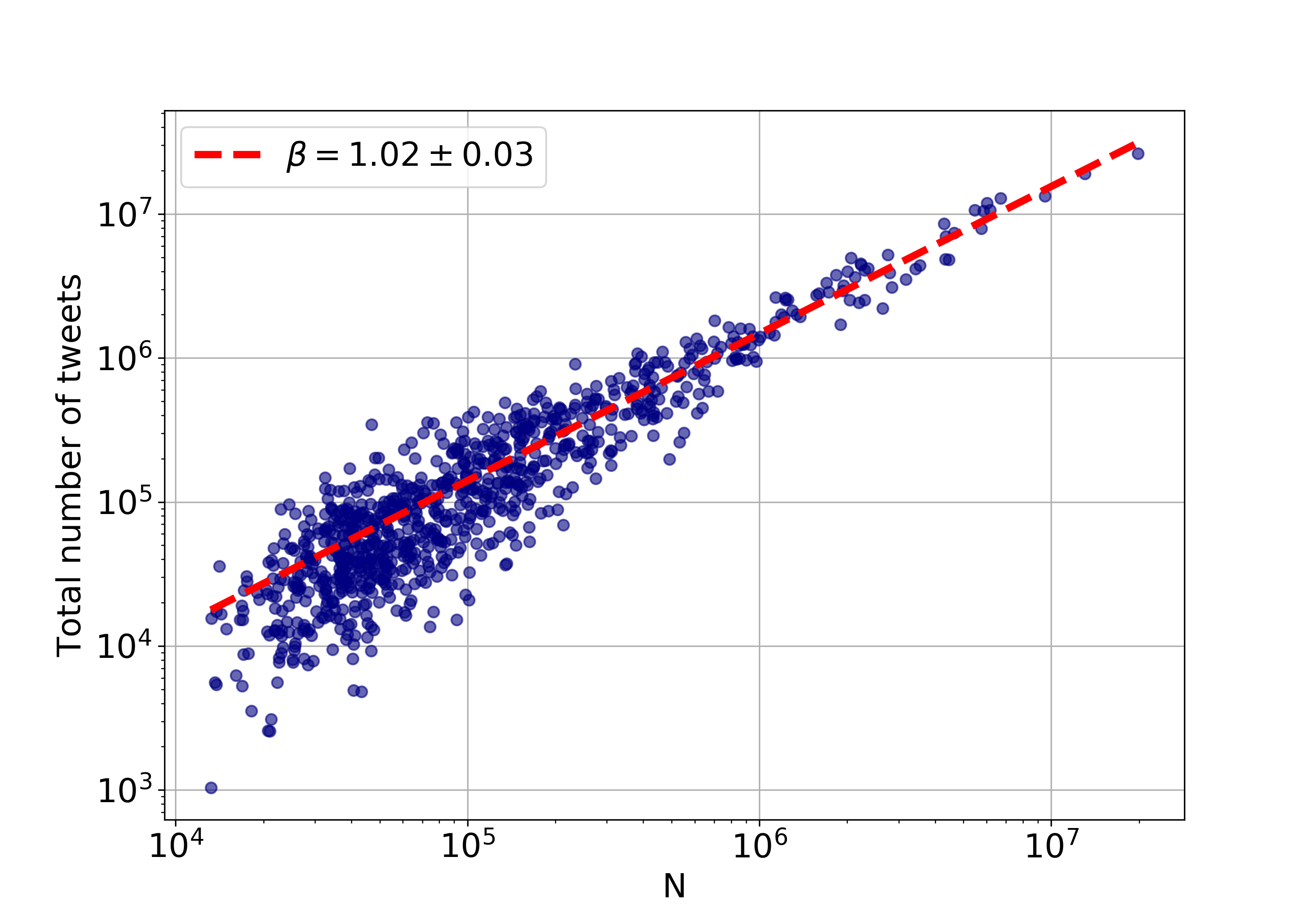}
\caption{Scaling of the total number of geolocated messages with city population. Each point represents an MSA, the fitted line is the best MLE fit for the Person Model of \cite{Leitao2016a}.}
\label{fig:tweetcnt}
\end{center}
\end{figure}

First, we checked how some aggregate metrics: the total number of users, the total number of individual words and the total number of tweets change with city size. Figures \ref{fig:usernum}, \ref{fig:wordcnt} and \ref{fig:tweetcnt} show the scaling relationship data on a log-log scale, and the result of the fitted model. In all cases, $\Delta BIC$ was greater than 6, which confirmed nonlinear scaling. The the total count of tweets and words both have a slightly superlinear exponents around 1.02. The deviation from the linear exponent may seem small, but in reality it means that for a tenfold increase in city size, the abundance of the quantity $Y$ measured increases by 5\%, which is already a significant change. The number of users scales sublinearly ($\beta=0.95\pm 0.01$) with the city population, though.

It has been shown in \cite{Schlapfer2014b} that total communication activity in human interaction networks grows superlinearly with city size. This is in line with our findings that the total number of tweets and the total word count scales superlinearly. However, the exponents are not as big as that of the number of calls or call volumes in the previously mentioned article ($\beta\in[1.08,1.14]$), which suggests that scaling exponents obtained from a mobile communication network cannot automatically be translated to a social network such as Twitter.

\subsection{Individual scaling of words}

\begin{figure}[h!]
\begin{center}
\subfloat[Sublinear scaling]{\includegraphics[width=0.3\textwidth]{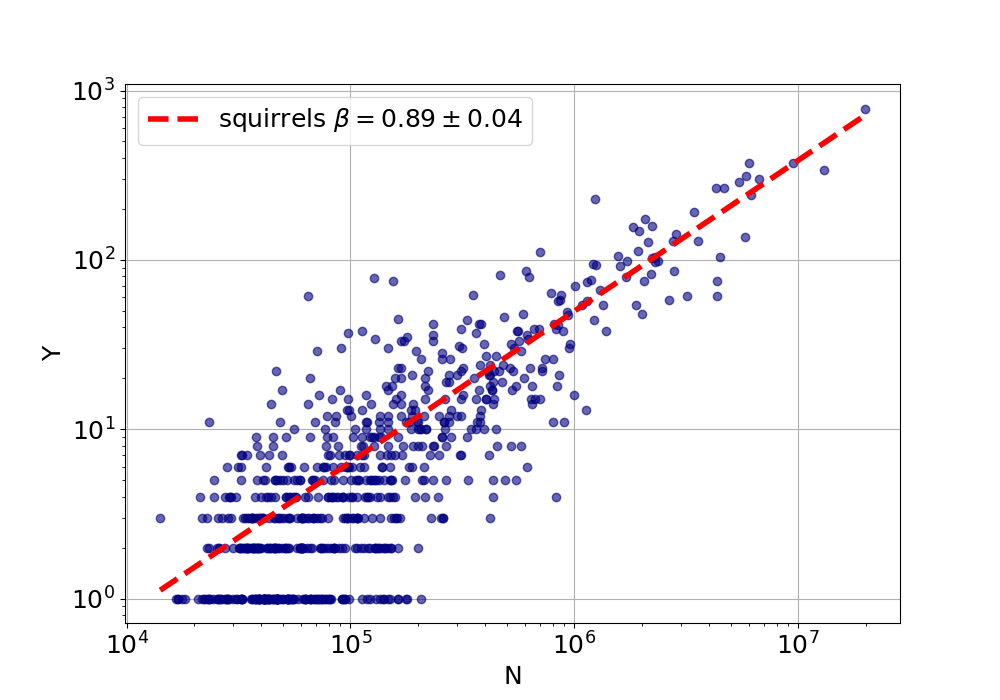}}
\subfloat[Linear scaling]{\includegraphics[width=0.3\textwidth]{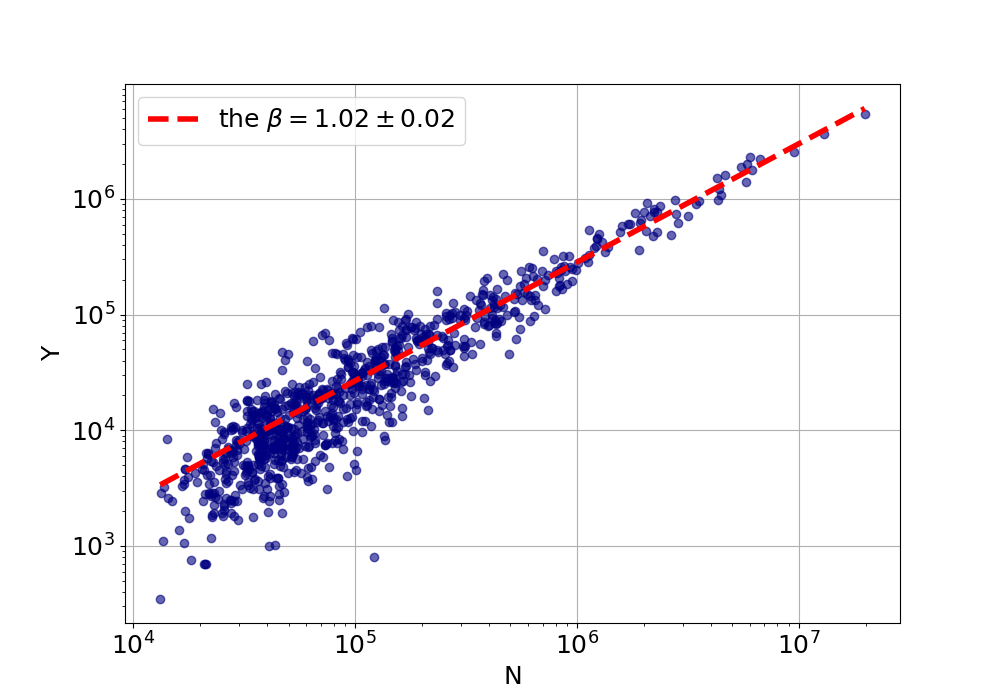}}
\subfloat[Superlinear scaling]{\includegraphics[width=0.3\textwidth]{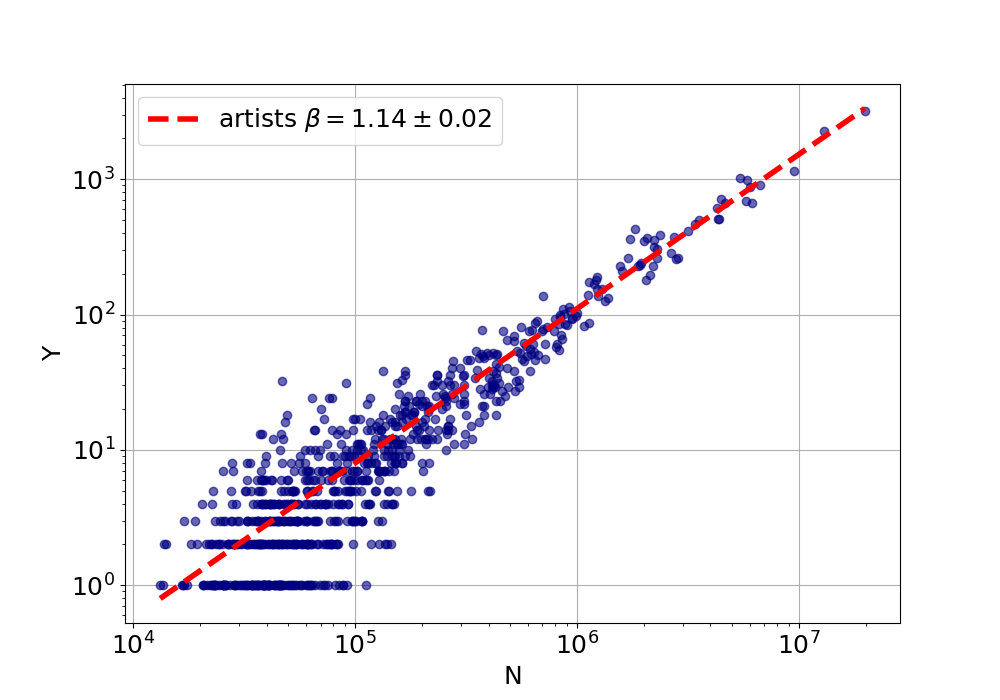}}
\caption{Three scaling relationships from the sublinear (a), linear (b), and superlinear (c) scaling regimes with the MLE fits explained in the Methods section.}
\label{fig:examples}
\end{center}
\end{figure}

\begin{figure}[h!]
\begin{center}
\includegraphics[width=0.6\textwidth]{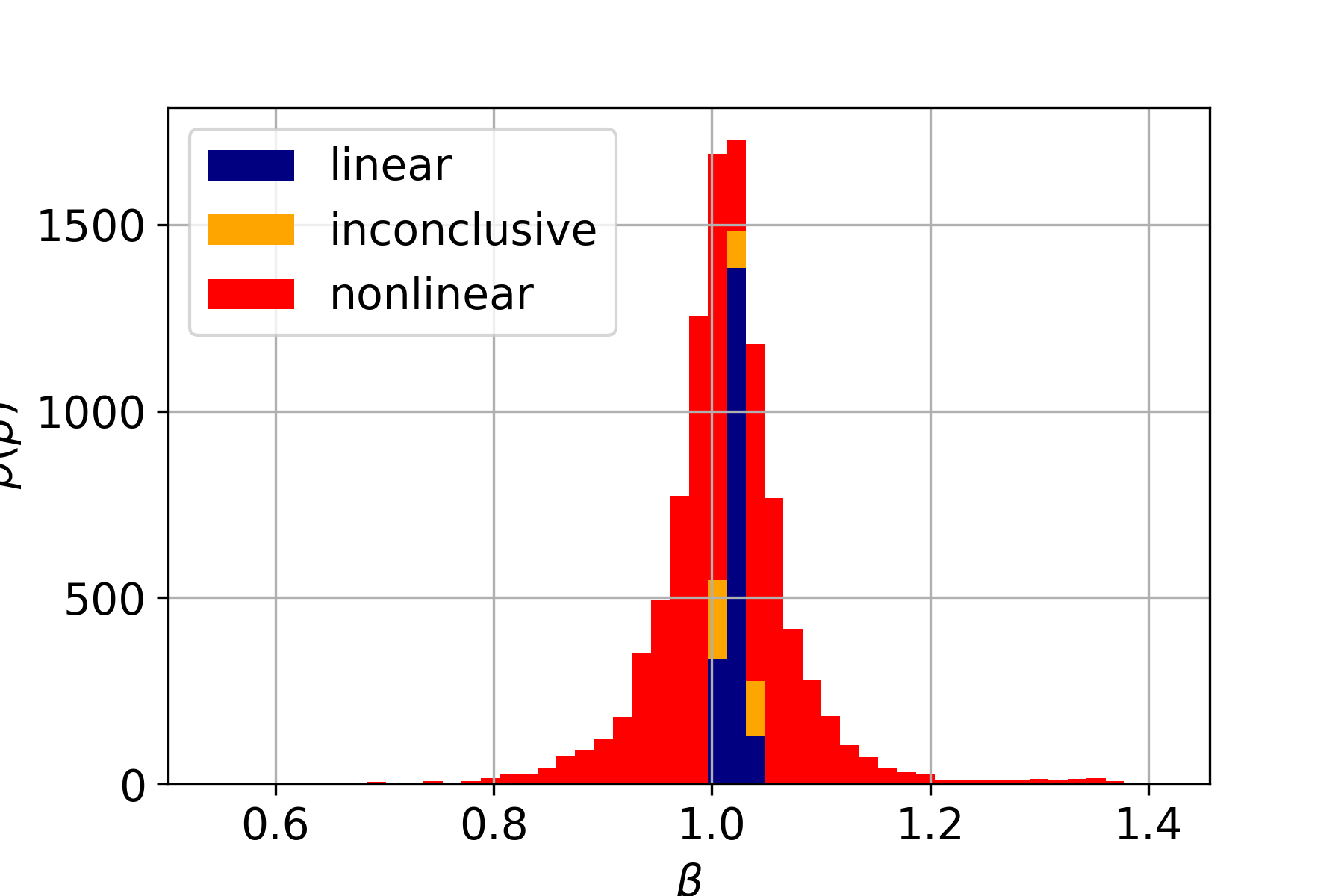}
\caption{Distribution of word exponents. Statistically significant deviations from the scaling of the total number of words are marked by color codes. The peak around 1.02 marks words that have an exponent around the exponent of the total number of words. The majority of words follow a superlinear or a sublinear scaling law. Note, that there can be multiple categories in one bin according to the $\Delta BIC$ of fits.}
\label{fig:expdistr}
\end{center}
\end{figure}

For the 11732 words that had at least 10000 occurrences in the dataset, we fitted scaling relationships using the Person Model. The distribution of the fitted exponents is visible in Figure~\ref{fig:expdistr}. There is a most probable exponent of approximately 1.02, which corresponds roughly to the scaling exponent of the overall word count. This is the exponent which we use as an alternative model for deciding nonlinearity, because a word that has a scaling law with the same exponent as the total number of words has the same relative frequency in all urban areas. The linear and inconclusive cases calculated from $\Delta BIC$ values are located around this maximum, as shown in different colors in Figure~\ref{fig:expdistr}. In this figure, linearly and nonlinearly classified fits might appear in the same exponent bin, because of the similarity in the fitted exponents, but a difference in the goodness of fit. Words with a smaller exponent, that are "sublinear" do not follow the text growth, thus, their relative frequency decreases as city size increases. Words with a greater exponent, that are "superlinear" will relatively be more prevalent in texts in bigger cities. There are slightly more words that scale sublinearly (5271, 57\% of the nonlinear words) than superlinearly (4011, 43\% of the nonlinear words). Three example fits from the three scaling regime are shown in Figure~\ref{fig:examples}.

\begin{table}
\begin{tabular}{p{\textwidth}}
\texttt{the you and that for this just lol like with have but get not your was all love what are when out know good now got can about one time day how they too shit want back need why she people right some see going today fuck will really her}
\end{tabular}
\caption{The top 50 words as ranked according to the $BIC$ values for a $\beta=1.0207$ fixed exponent Person Model. These are the words that correspond most to the scaling of the overall word volume, thus, they are the words that appear most homogeneously in the texts of all urban areas.}
\label{table:toplinear}
\end{table}

We sorted the words falling into the "linear" scaling category according to their $BIC$ values showing the goodness of fit for the fixed $\beta$ model. The first 50 words in Table~\ref{table:toplinear} according to this ranking are some of the most common words of the English language, apart from some swearwords and abbreviations (e.g. lol) that are typical for Twitter language \cite{Bokanyi2016}. These are the words that are most homogeneously present in the text of all urban areas.

\begin{table}
\begin{tabular}{p{\textwidth}}
sublinear:\\
\texttt{flood severe thunderstorm warning statement april lsu february bama ole unc shxt beside deer shelby kentucky ian fishing dynasty dorm freeze nigha carolina roomie walmart december january tornado gotti mountains mite wind kelsey campus exams mart roommates frat mud roads lmbo biology duke logan roommate ruzzle exam pinterest brooke bahaha slowly further mam hunting bahahaha thanking dang dwn hush softball bailey haley porch rec gates yuu november memphis marshall haven storms ncaa cody renee tanning oomf heck paige nosey casino southern muh bre lab tub truck cowboy jeep seth messy lawd layin tourney trashy puke library gah lake tweeps rae semester wreck johns bonfire studying until quit state gotcha anatomy prolly knw eagle wrk lifting flag lastnight courtney awhile tweetin bend ann abby march douche snuggle fog bracket hannah bedtime golf sittin gosh lynn whiskey nerves rain road town fixing hut whatcha drinkin driveway damnit country moore riley lyin duck}\\
\\
superlinear:\\
\texttt{hoy gracias por para feliz con cuando que siempre verdad algo donde amor ver tiempo mejor semana estas alguien bien jajaja mas del todo jajajaja vez tus ama tengo vamos porque buenos eres linda muy quiero puedo hola las mucho nada sabes mañana amo soy les tambien vas dormir buenas amigo hay madrid mis bueno gusta brunch mal jaja uno flight familia dos cara delayed landed dice casa amigos loco grande papi fin traffic tix com lounge puerto heights brazil rico deja gate madre solo pls luis plane event international bon bella oscar sin mil damm ily mon studio maria carlos lmfao italian das film thx omw peep era salon omfg van jose london sushi blocks security vip mah ilysm hookah fitness cos ariana fashion via park jenny performing pronto artists stadium kanye restaurant awk melissa market danny ale booked leo inspired connect rft fab culture artist demi blasting design}
\end{tabular}
\caption{The most sublinearly ($0.54<\beta<0.93$) or superlinearly ($1.13<\beta<1.41$) scaling words out of the 5000 most frequent words with small bootstrapped error $\Delta \beta <0.1$. Sublinear words are sorted in an ascending, superlinear words in a descending order with respect to $\beta$.}
\label{table:wordlist}
\end{table}

From the first 5000 words according to word rank by occurrence, the most sublinearly and superlinearly scaling words can be seen in Table~\ref{table:wordlist}. Their exponent differs significantly from that of the total word count, and their meaning can usually be linked to the exponent range qualitatively. The sublinearly scaling words mostly correspond to weather services reporting (flood 0.54, thunderstorm 0.61, wind 0.85), some certain slang and swearword forms (shxt 0.81, dang 0.88, damnit 0.93), outdoor-related activities (fishing 0.82, deer 0.81, truck 0.90, hunting 0.87) and certain companies (walmart 0.83). There is a longer tail in the range of superlinearly scaling words than in the sublinear regime in Figure \ref{fig:expdistr}. This tail corresponds to Spanish words (gracias 1.41, por 1.40, para 1.39 etc.), that could not be separated from the English text, since the shortness of tweets make automated language detection very noisy. Apart from the Spanish words, again some special slang or swearwords (deadass 1.52, thx 1.16, lmfao 1.17, omfg 1.16), flight-reporting (flight 1.25, delayed 1.24 etc.) and lifestyle-related words (fitness 1.15, fashion 1.15, restaurant 1.14, traffic 1.22) dominate this end of the distribution.

Thus, when compared to the slightly nonlinear scaling of total amount of words, not all words follow the growth homogeneously with this same exponent. Though a significant amount remains in the linear or inconclusive range according to the statistical model test, most words are sensitive to city size and exhibit a super- or sublinear scaling. Those that fit the linear model the best, correspond to a kind of 'core-Twitter' vocabulary, which has a lot in common with the most common words of the English language, but also shows some Twitter-specific elements. A visible group of words that are amongst the most super- or sublinearly scaling words are related to the abundance or lack of the elements of urban lifestyle (e.g. deer, fitness). Thus, the imprint of the physical environment appears in a quantifiable way in the growths of word occurrences as a function of urban populations. Swearwords and slang, that are quite prevalent in this type of corpus \cite{Gauthier2015TextHabits,Wang2014CursingTwitter},  appear at both ends of the regime that suggests that some specific forms of swearing disappear with urbanization, but the share of overall swearing on Twitter grows with city size. The peak consisting of Spanish words at the superlinear end of the exponent distribution marks the stronger presence of the biggest non-English speaking ethnicity in bigger urban areas. This is confirmed by fitting the scaling relationship to the Hispanic or Latino population \cite{RankingCenter} of the MSA areas ($\beta=1.31\pm 0.14$, see SI), which despite the large error, is very superlinear. 

\subsection{Zipf's law on Twitter}

\begin{figure}[h!]
\begin{center}
\includegraphics[width=0.6\textwidth]{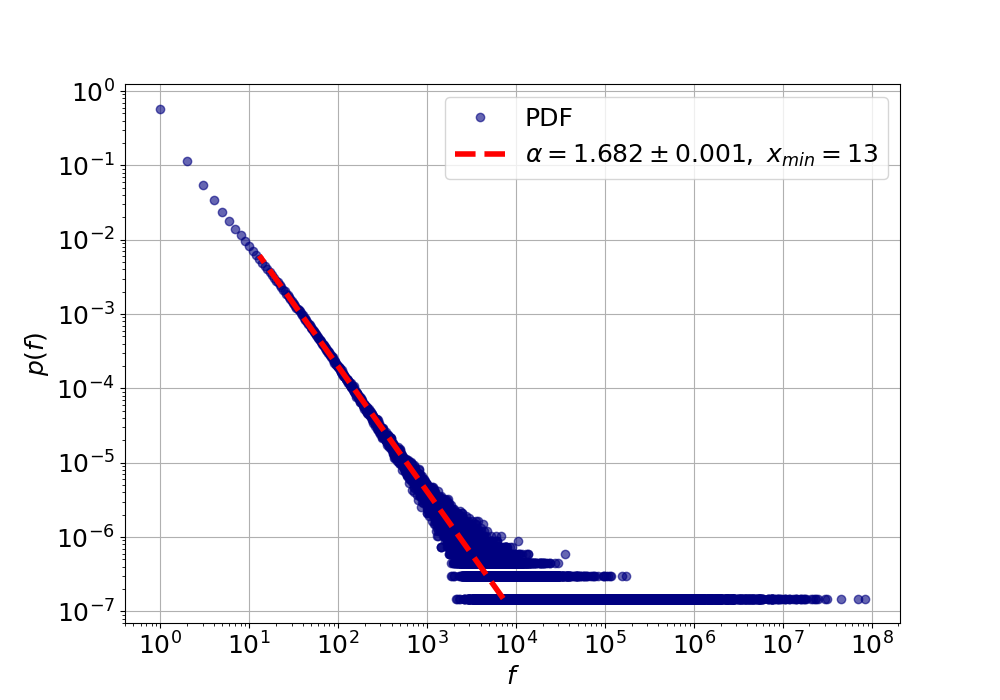}
\caption{Probability distribution of word frequencies in the overall corpus and power-law fitted by the \texttt{powerlaw} package.}
\label{fig:zipf}
\end{center}
\end{figure}

\begin{figure}[h!]
\begin{center}
\includegraphics[width=0.6\textwidth]{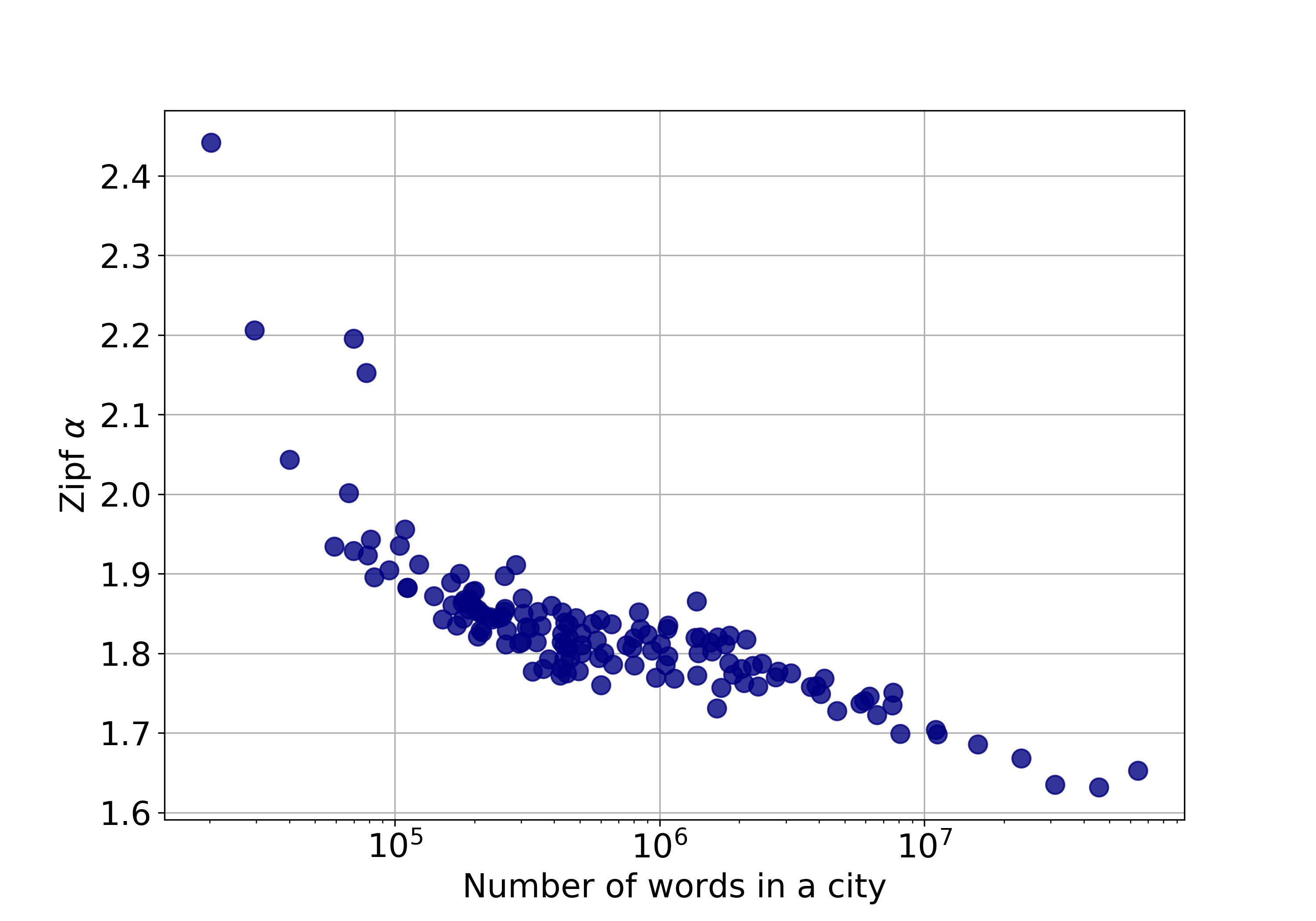}
\caption{Dependency of the Zipf exponent on city population. The exponent decreases as the number of words in a city grows.}
\label{fig:zipfcity}
\end{center}
\end{figure}

Figure~\ref{fig:zipf} shows the distribution of word counts in the overall corpus. The power-law fit gave a minimum count $x_{min}=13$, and an exponent $\alpha=1.682\pm0.001$. To check whether this law depends on city size, we fitted the same distribution for the individual cities, and according to Figure~\ref{fig:zipfcity}, the exponent gradually decreases with city size, that is, it decreases with the length of the text. 

That the relative frequency of some words changes with city size means that the frequency of words versus their rank, Zipf's law, can vary from metropolitan area to metropolitan area. We obtained that the exponent of Zipf's law depends on city size, namely that the exponent decreases as text size increases. It means that with the growth of a city, rarer words tend to appear in greater numbers. The values obtained for the Zipf exponent are in line with the theoretical bounds 1.6-2.4 of \cite{FerreriCancho2005TheLanguage}. In the communication efficiency framework \cite{FerreriCancho2005TheLanguage,FerreriCancho2003}, decreasing $\beta$ can be understood as decreased communication efficiency due to the increased number of different tokens, that requires more effort in the process of understanding from the reader. Using more specific words can also be a result of the 140 character limit, that was the maximum length of a tweet at the time of the data collection, and it may be a similar effect to that of texting \cite{Crystal2008Texting}. This suggests that the carrying medium has a huge impact on the exact values of the parameters of linguistic laws.

The Zipf exponent measured in the overall corpus is also much lower than the $\beta = 2$ from the original law \cite{Zipf1932SelectedLanguage}. We do not observe the second power-law regime either, as suggested by \cite{Montemurro2001} and \cite{FerreriCancho2001}. Because most observations so far hold only for books or corpora that contain longer texts than tweets, our results suggest that the nature of communication, in our case Twitter itself affects the parameters of linguistic laws.

\subsection{Vocabulary size change}

\begin{figure}[h!]
\begin{center}
\includegraphics[width=0.6\textwidth]{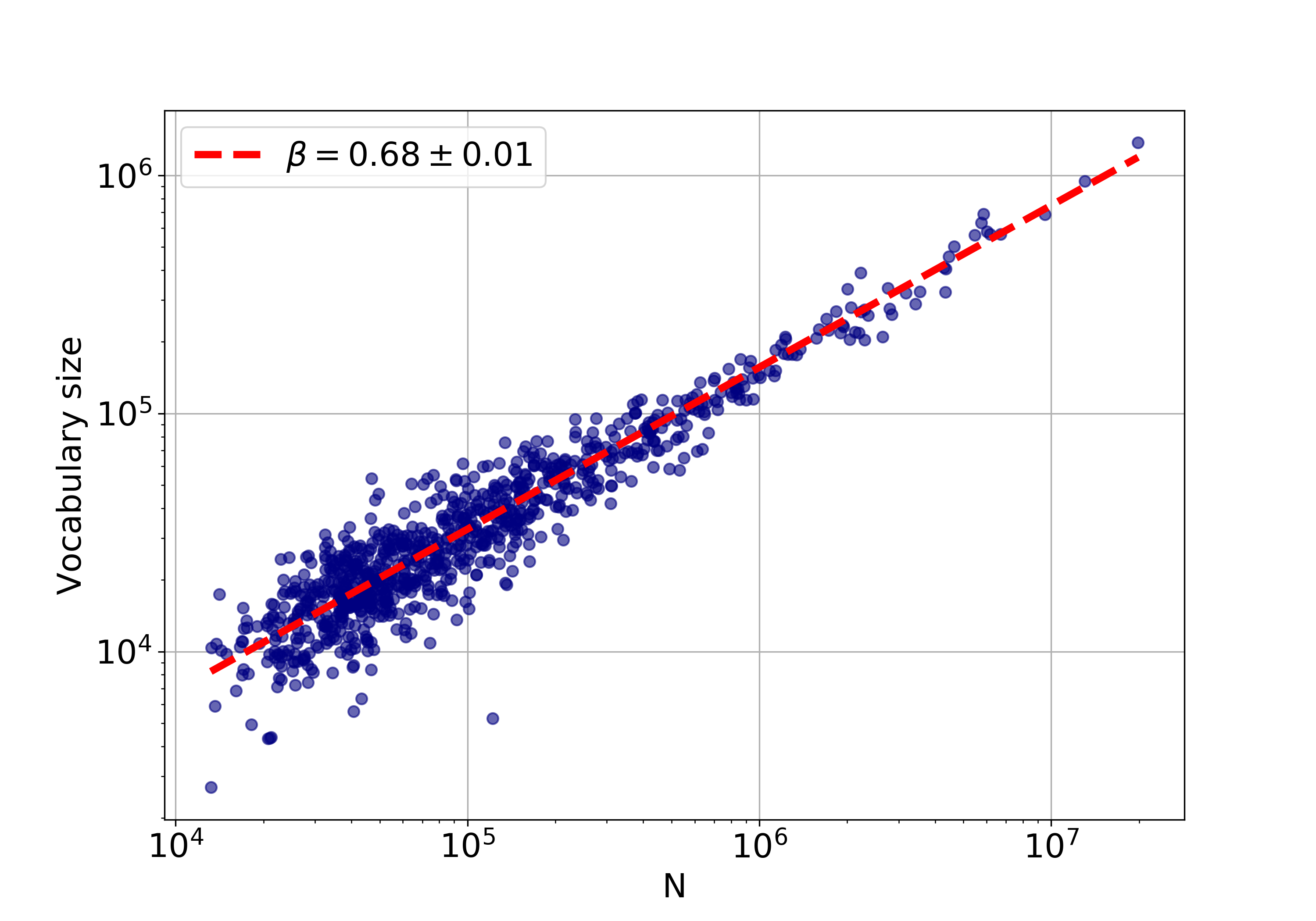}
\caption{Scaling of the total number of distinct words with city population. Each point represents an MSA, the fitted line is the best MLE fit for the Person Model of \cite{Leitao2016a}.}
\label{fig:vocabcnt}
\end{center}
\end{figure}

Figure~\ref{fig:vocabcnt} shows the vocabulary size as a function of the metropolitan area population, and the power-law fit. It shows that in contrary to the previous aggregate metrics, the vocabulary size grows very sublinearly ($\beta=0.68$) with the city size. This relationship can also be translated to the dependency on the total word count, which would give a $\beta=0.68/1.02=0.67$, another sublinear scaling.

The decrease in $\beta$ for bigger cities (or bigger Twitter corpora) suggesting a decreasing number of words with lower frequencies is thus confirmed. There is evidence, that as languages grow, there is a decreasing marginal need for new words \cite{Petersen2012}. In this sense, the decelerated extension of the vocabulary in bigger cities can also be regarded as language growth. 

\section{Conclusion}

In this paper, we investigated the scaling relations in citywise Twitter corpora coming from the Metropolitan and Micropolitan Statstical Areas of the United States. We could observe a slightly superlinear scaling decreasing with the city population for the total volume of the tweets and words created in a city. When observing the scaling of individual words, we found that a certain core vocabulary follows the scaling relationship of that of the bulk text, but most words are sensitive to city size, and their frequencies either increase at a higher or a lower rate with city size than that of the total word volume. At both ends of the spectrum, the meaning of the most superlinearly or most sublinearly scaling words is representative of their exponent. We also examined the increase in the number of words with city size, which has an exponent in the sublinear range. This shows that there is a decreasing amount of new words introduced in larger Twitter corpora.

\clearpage

\setlength{\parskip}{0em}

\subsubsection*{Data availability}

Owing to Twitter's policy we cannot publicly share the original dataset used in this analysis. However, aggregated results from which all calculations can be recreated are available in at \url{http://bokae.web.elte.hu/papers/2018/word_scaling} and from the Dryad Digital Repository: \url{https://doi.org/10.5061/dryad.824f24t}, with the review link \url{ https://datadryad.org/review?doi=doi:10.5061/dryad.824f24t}.

\subsubsection*{Competing interests}

The authors declare no competing interests.

\subsubsection*{Author contributions}

E.B. and G.V. designed the study, E.B. and D.K. analyzed the data, E.B., D.K. and G.V. synthetized the results, E.B. and D.K. wrote the manuscript. All authors gave final approval for publication and agree to be held accountable for the work performed therein.

\subsubsection*{Funding}

The authors thank the support of the National Research, Development and Innovation Office of Hungary (grant no. KH125280).

\subsubsection*{Research Ethics}
We were not required to complete an ethical assessment prior to conducting our research.

\subsubsection*{Animal Ethics}
We were not required to complete an ethical assessment prior to conducting our research.

\subsubsection*{Permission to carry out fieldwork}

No permissions were required prior to conducting our research.

\subsubsection*{Acknowledgements}

Not applicable.

\clearpage

\bibliographystyle{unsrt}
\bibliography{biblio}

\end{document}